\documentclass[preprint]{revtex4-1}

\usepackage[UKenglish]{babel}
\usepackage[latin9]{inputenc}
\usepackage[T1]{fontenc}
\usepackage{graphicx} \usepackage{graphics}
\usepackage{soul} 

	\newcommand{\LSCO}{La$_{1.85}$Sr$_{0.15}$CuO$_{4}$}
	\newcommand{\LxSCO}{La$_{2-x}$Sr$_{x}$CuO$_{4}$}
	\newcommand{\STO}{SrTiO$_3$}
	\newcommand{\CTO}{CaTiO$_3$}

	\newcommand{\Tc}{$T_{{\rm c}}$}
	\newcommand{\Ba}{BaFe$_{1.8}$Co$_{0.2}$As$_2$}
	\newcommand{\Bax}{BaFe$_{2-x}$Co$_x$As$_2$}
	\newcommand{\x}{La$_{2-x}$Sr$_{x}$CuO$_{4}$}

\begin{document}

\author{S. Trommler} 
\altaffiliation{Dresden University of Technology, Department of Physics, Institute for Physics of Solids, 01062 Dresden}
\author{R. Hühne}
\author{K. Iida}
\author{P. Pahlke}
\author{S. Haindl}
\author{L. Schultz} 
\altaffiliation{Dresden University of Technology, Department of Physics, Institute for Physics of Solids, 01062 Dresden}
\author{B. Holzapfel  }
\altaffiliation{Dresden University of Technology, Department of Physics, Institute for Physics of Solids, 01062 Dresden, Germany}
\affiliation{IFW Dresden, P.O. Box 270116, 01171 Dresden, Germany, electronic mail: s.trommler@ifw-dresden.de}

\title{Reversible shift in the superconducting transition for \LSCO\ and \Ba\ using piezoelectric substrates}

\begin{abstract}

The use of piezoelectric substrates enables a dynamic observation of strain dependent properties of functional materials. Based on studies with \LSCO\ we extended this approach to the iron arsenic superconductors represented by \Bax\ to investigate strain driven changes in detail. We demonstrate that epitaxial thin films can be prepared on (001) Pb(Mg$_{1/3}$Nb$_{2/3}$)$_{0.72}$Ti$_{0.28}$O$_3$ substrates using pulsed laser deposition. The structural as well as the electric properties of the grown films were characterized in detail. A reversible shift of the superconducting transition of 0.44 K for \LSCO\ and 0.2 K for \Ba\ was observed applying a biaxial strain of 0.022\%\ and 0.017\%\, respectively.


\end{abstract}


\maketitle


The application of pressure has a significant influence on the physical properties of functional materials. Detailed experiments are required to enable a deeper understanding in physics of these materials especially the sensitive interplay between structural parameters like bonding length or angle and the electronic properties.
Existing studies on bulk materials predominantly cuprate high--temperature superconductors using hydrostatic pressure 
demonstrate that compressive pressure increases the superconducting transition temperature, \Tc, for most materials. Recent studies also reveal a strong influence of pressure on superconductivity for iron based superconductors resulting in a pressure dependent superconducting dome in the electronic phase diagram similar to doping \cite{Kim09, Oka08}.
It should be noted that this shift in \Tc\ is highly anisotropic regarding strain along different crystallographic axes. For some oxides like \x\ it is known that the resulting effect is partially neutralized for hydrostatic pressure \cite{Bud92, Gug94, Che91, noh95, Che91}.  

Therefore, in the last decade also the application of biaxial strain attracted increasing attention, especially for the model system \LxSCO\ \cite{sat04, sat08}. Similarly, recent studies on biaxial strained iron based superconductors like BaFe$_{1.8}$Co$_{0.2}$As$_2$ and FeSe$_{0.5}$Te$_{0.5}$ also revealed that compressive biaxial strain enhances \Tc \cite{Iid09, Bel10}.
Typically, epitaxial thin films are prepared on various single crystalline substrates with a different lattice mismatch between substrate and film inducing a biaxial tensile or compressive strain. However, this approach is often restricted to very thin films due to a limited layer thickness for coherent strained growth. A large misfit typically leads to partial relaxation of the lattice and, therefore, to the implementation of lattice defects. In this case it is difficult to correlate the applied strain with superconducting properties directly, as the preparation conditions and the resulting microstructure may severely affect the latter. 

An alternative approach to static pressure experiments is the preparation of superconducting films on single crystalline piezoelectric substrates. Using the inverse piezoelectric effect the applied strain can be changed  continuously and reversibly by an electric field. 
This approach offers the unique opportunity to investigate the strain dependent properties on one and the same sample as shown already for ferromagnetic oxides \cite{thi07,dor07, rat08, Her09}. Recently we reported on the epitaxial growth of superconducting YBa$_2$Cu$_3$O$_{7-\delta}$ and \LSCO\ (LSCO) thin films on pseudocubic (001) Pb(Mg$_{1/3}$Nb$_{2/3}$)$_{0.72}$Ti$_{0.28}$O$_3$ (PMN-PT) substrates \cite{Huh08, Tro09}. In this letter we extend this approach to \Ba\ (Ba-122) thin films and report on a reversible shift in \Tc\ with applied strain.



\begin{figure}[ht]
\begin{center}\includegraphics[width=16.5cm]{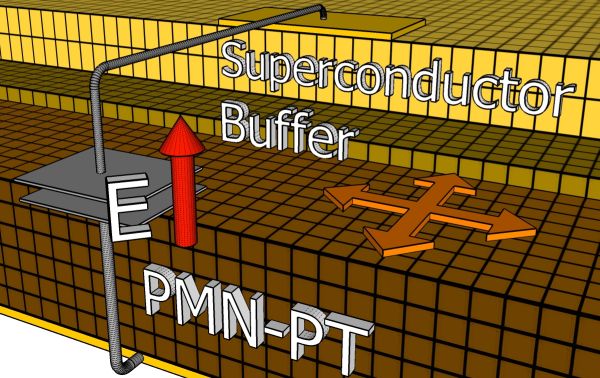}\end{center}
\caption{Schematic film architecture. The bottom contacts are sputtered NiCr/Au, whereas the top electrodes are deposited gold.} \label{scetch}
\end{figure}

For the sample preparation on the PMN-PT we used a standard pulsed laser deposition (PLD) setup equipped with a Lambda Physiks LPX 305 KrF laser and stoichiometric targets. To reduce the lattice mismatch between PMN-PT ($a$=4.02 \AA) and the superconducting film we deposited smooth 20 nm thick buffer layers of either \STO\ (STO) ($a$=3.905 \AA) or \CTO\ (CTO) ($a$=3.82 \AA) \cite{Bil08}. The buffer layers as well as the 300 nm thick LSCO films are prepared in 0.3 mbar oxygen atmosphere at substrate temperatures of 650°C--700°C using off--axis deposition \cite{rat08, hol92}. Films prepared under this condition typically exhibit a very smooth surface and droplet--free growth \cite{hol92}. Subsequently, the films were cooled down in 0.4 bar oxygen atmosphere. A detailed description of the LSCO preparation as well as structural analysis can be found in our previous publication \cite{Tro09}.  A scheme of the film architecture is given in fig.\ref{scetch}.

For the preparation of Ba-122 we used STO buffered PMN-PT prepared by off--axis--PLD as described above. Subsequently the substrate was transferred to an ultra high vacuum system with a base pressure of 10$^{-9}$ mbar where the Ba-122 was deposited at 650°C using on--axis--PLD. A detailed description of the film preparation can be found in Iida et al. \cite{Iid09,Iid10}. 


The superconducting properties were characterized in a Quantum Design Physical Properties Measurement System (PPMS) using a four probe technique. For the evaluation of the transition temperature a 50\%\ resistance criterion is used.
To confirm epitaxial growth and to study the structural properties
standard x-ray diffraction (XRD) in Bragg--Brentano geometry, pole figure measurements and reciprocal space mapping (RSM) were performed using a Phillips XPert MRD Diffractometer with Cu K$_{\alpha}$ radiation. X--ray reflectivity was used to determine the layer thickness and the roughness of the buffer layers.

\begin{figure}[ht]
\begin{center}\includegraphics[width=16.5cm]{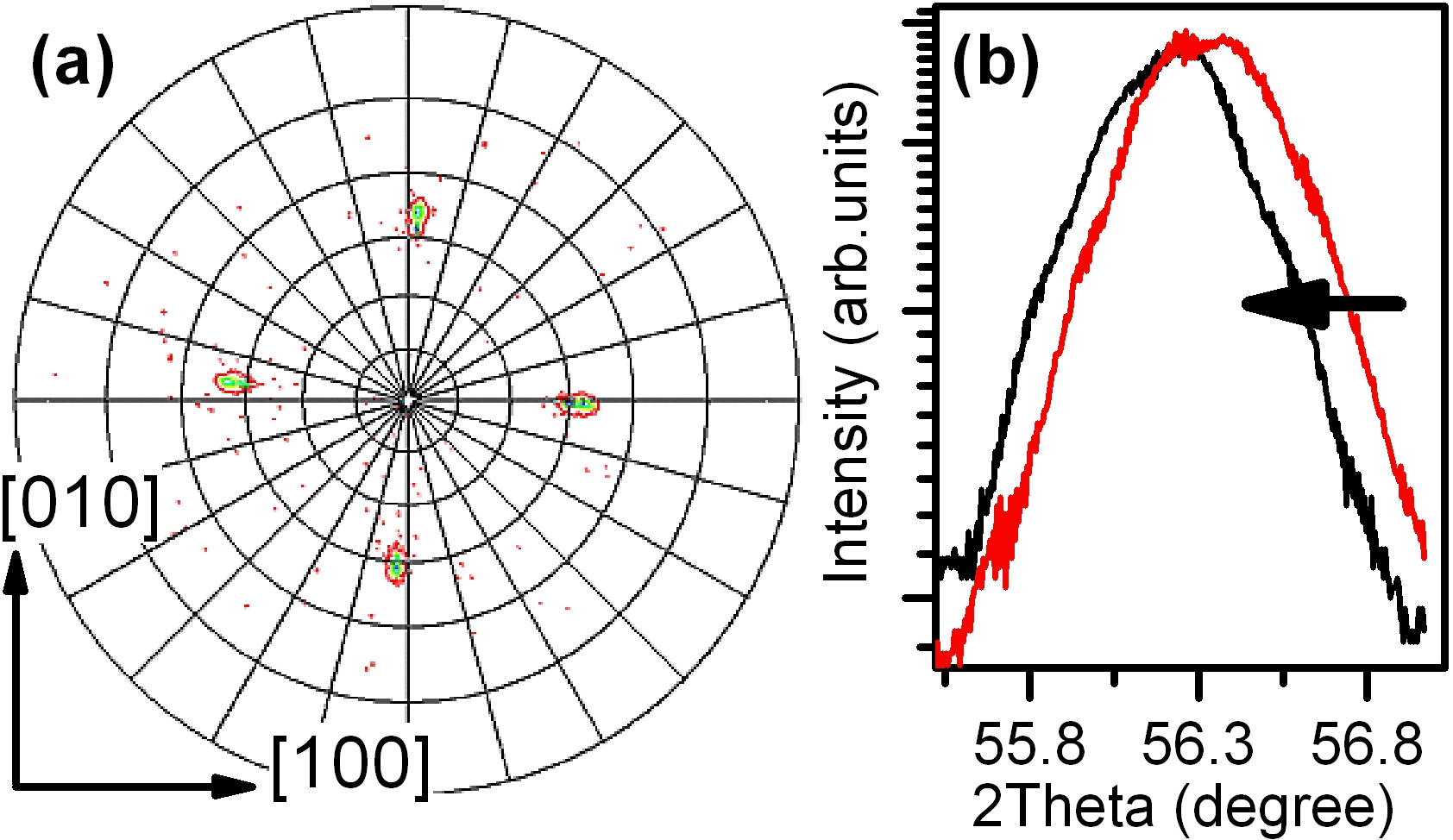}\end{center}
\caption{ (a) Pole figure of the Ba-122 (103) reflection. The intensity is scaled quadratic, where maximum intensity corresponds to 650 cps; (b) the \Ba\ (008) peak is shifted to lower angles when compressive in--plane strain is applied.} \label{XRD_122}
\end{figure}

We achieved a {\it c}--axis oriented growth and cube on cube epitaxy for both, buffer layer and LSCO \cite{Tro09}. Also for Ba-122 the pole figure of the (103) peak, given in fig.\ref{XRD_122}(a), proves perfect cube on cube epitaxy without any misorientation since the \Ba\ peaks are oriented parallel the substrate [100] directions.
The superconducting transition of LSCO at 17.5 K on CTO buffered PMN-PT is slightly smaller compared to 18.5 K on STO buffered PMN-PT. However, the CTO buffered system was used for further investigations due to reduced affinity to crack during dynamic strain measurements.

In the case of \Ba\ the STO buffered films exhibit a \Tc\ of 14 K which is significantly reduced compared to 23 K for films prepared on bare STO \cite{Iid09}. Part of this reduction is attributed to the poorer crystalline quality of the PMN-PT substrate compared to STO and the much larger transition width as we use a 50\% criterion.


In the first step it is necessary to ensure the transfer of strain into the superconducting layer. Detailed investigations by Bilani et al. showed that the strain is transferred from the PMN-PT to the STO buffer \cite{Bil08}. We performed additional high resolution XRD and RSM to verify the strain transfer to the superconducting layer. An example is given in fig.\ref{XRD_122}(b) for the Ba-122 (008) peak without and with 16.6 kV/cm applied field. 

The change of the lattice parameters in PMN-PT single crystals at room temperature in dependence of the applied electric field is well investigated \cite{Par97}, however, there is no data available for lower temperatures. Nevertheless, the knowledge of the low temperature behavior is essential to correlate the strain with the change in the superconducting properties.

\begin{figure}[ht]
\begin{center}\includegraphics[width=16.5cm]{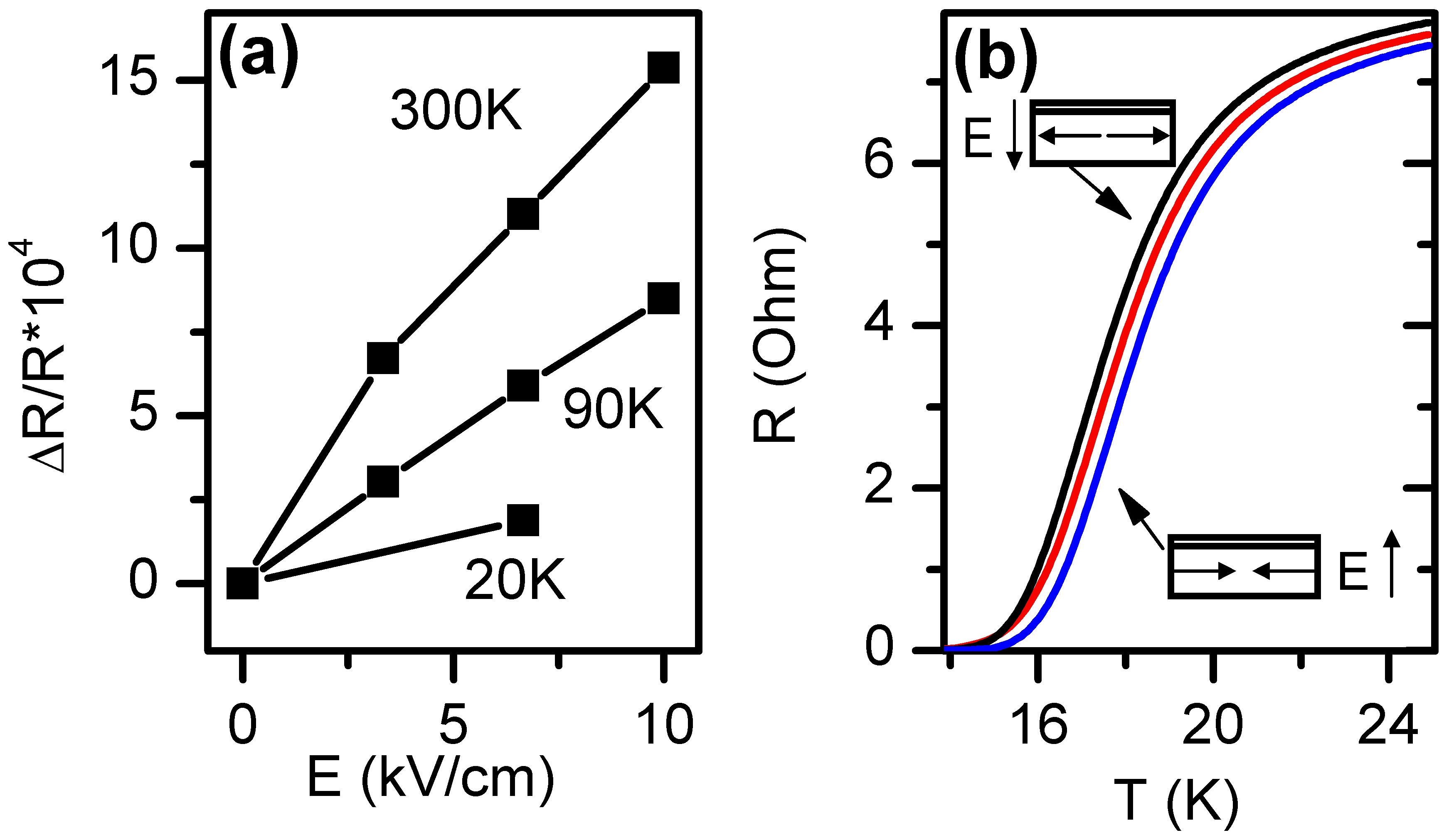}\end{center}
\caption{ (a) Relative change of the resistivity of a platinum strain gauge in dependence of the field for three different temperatures; (b) shift of transition temperature with applied field for LSCO for E=-10 kV/cm (black), E=0 kV/cm (red) and E=10 kV/cm (blue).} \label{Tc_LSCO}
\end{figure}

To gauge the magnitude of strain at lower temperatures we deposited a thin meander-shaped platinum wire at room temperature directly on CTO buffered PMN-PT. The resistivity of the wire correlates to the biaxial strain due to a change of the wire geometry.
The  change of the relative resistivity with the applied electric field is given in fig.\ref{Tc_LSCO}(a) for three different temperatures. 
This data reveals a strong reduction of the strain at constant electric field with decreasing temperature. 
Compared to room temperature ($\epsilon_a$=0.12\% at 10 kV/cm) we achieve half the value at 90 K and less than 20\% at 20 K. The biaxial in--plane strain,  $\epsilon_a$, is defined as $(a_{0}$-$a_{strained})$/$a_{0}$, where $a_{0}$ is the unstrained in--plane lattice parameter.


To check the suitability of our approach we used the well known model system LSCO.
Applying an electric field of $E$=10 kV/cm at 20 K which corresponds to $\epsilon_a$=0.022\% we achieved a reversible shift of the superconducting transition temperature of 0.4 K (fig.\ref{Tc_LSCO}(b)). We compared this shift to available literature data using a simple equation of the strain dependent \Tc\ for an orthorhombic unit cell, where \Tc (0) denotes the superconducting transition temperature of the unstrained film:  

\begin{equation}
T_{\mbox{c}} = T_{\mbox{c}}(0) + \frac{\delta T_{\mbox{c}}}{\delta \epsilon_a} \epsilon_a + \frac{\delta T_{\mbox{c}}}{\delta \epsilon_b} \epsilon_b + \frac{\delta T_{\mbox{c}}}{\delta \epsilon_c} \epsilon_c
\end{equation}

The values for the derivatives are well investigated for LSCO \cite{Gug94}. Due to the biaxial strain $\epsilon_a$ equals $\epsilon_b$. Taking the correlation of $\epsilon_c$ and $\epsilon_a$ from statically strained LSCO films \cite{sat08} we can replace $\epsilon_c$ in the the out--of--plane term and finally summarize equation(1) to:

\begin{equation}
T_{\mbox{c}} = T_{\mbox{c}}(0) + \beta  \epsilon_a, 
\end{equation}

where $\beta$=2000 K. This results in a theoretical change of the transition temperature of 0.44 K for $\epsilon_a$=0.022\%, which is in good agreement with our experimental results (fig.\ref{Tc_LSCO}(b)).



\begin{figure}[ht]
\begin{center}\includegraphics[width=16.5cm]{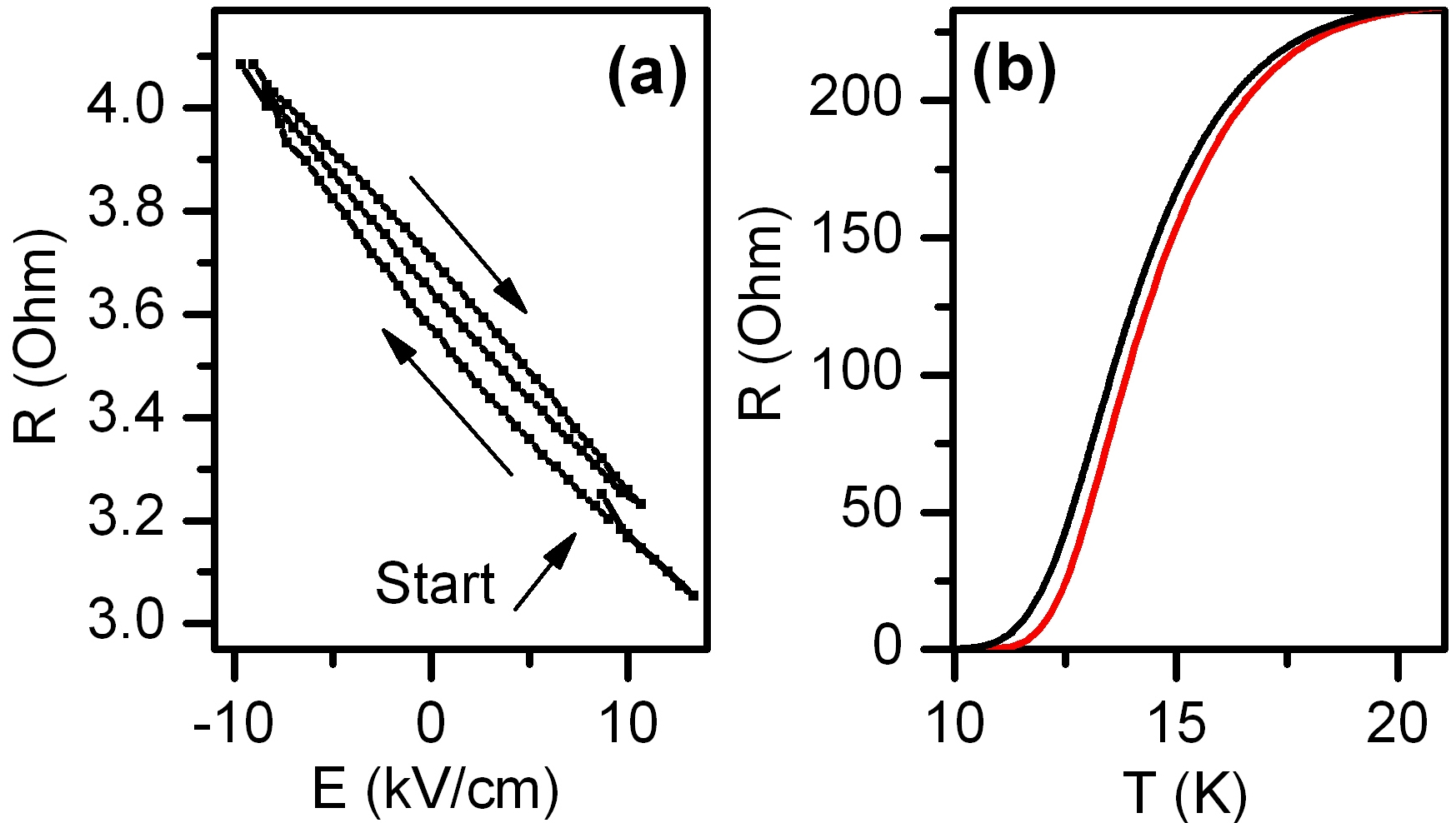}\end{center}
\caption{(a) Reversible change of the resistivity over the applied field at a fixed temperature of 18 K; (b) superconducting transition temperature of Ba-122 at E=-6.6 kV/cm (black) and E=6.6 kV/cm (red).} \label{reversibility}
\end{figure}

We checked the reversibility of the applied strain and the relaxation time of the PMN-PT at low temperatures, which is the time a piezoelectric material needs to reach the equilibrium strained state. We measured the resistivity depending on the applied electric field at a fixed temperature within the superconducting transition at 18 K. At this point the slope is very steep enabling the detection of minor changes in resistivity, when the transition curve is shifted. Starting at E=8.66 kV/cm we determined the resistivity by varying the field in steps of 0.66 kV/cm. By successive change of the electric field a reversible change in resistivity was obtained (fig.\ref{reversibility}(a)).
The minor deviation from linear behavior we attribute to the fact, that the time between field change and data acquisition was too less to reach the equilibrium strain state. To characterize the relaxation time we reverse the polarity of the electric field starting at 13.3 kV/cm within 5 seconds and subsequently measured the resistivity depending on the time.
Choosing a criterion for the equilibrium of less than 1\% resistance change per hour, the equlibrium is reached after 30 minutes.

Applying an electric field to an STO buffered Ba-122 film we observed a shift of the superconducting transition of 0.2 K for $\epsilon_a$=0.017\% (fig.\ref{reversibility}(b)) corresponding to $\beta$=1700 K. 
We compared the data with the results on statically epitaxial strained Ba-122 thin films where compressive strain results in different {\it c}/{\it a} ratios.
There a strain of $\epsilon_a$=1.2\%\ was achieved resulting in a shift of the critical temperature of about 8 K \cite{Iid09}. The corresponding $\beta$=670 K is less than half the value, we achieve with the dynamic approach. Analyzing this difference one has to take into account that equation(1) is only valid for small strain effects. In addition hydrostatic pressure experiments reveal a non linear change in \Tc\ with pressure \cite{Kim09}. we expect a similar behavior for biaxial strain.




In summary, we demonstrated the suitability of the inverse piezoelectric effect for the dynamical investigation of stain dependent superconducting properties. We observed a significant change in the superconducting transition temperature for both, LSCO and Ba-122 thin films. We conclude that compressive biaxial strain enhances the critical temperature for Ba-122 similar to cuprates like LSCO.

\begin{acknowledgments}
This work was partially supported by the German Research Foundation.
\end{acknowledgments}

\bibliographystyle{apsrev}

\newpage


\end{document}